\documentclass[preprint,3p,times,twocolumn]{elsarticle}

\usepackage{latexsym} 
\usepackage{graphicx} 
\usepackage{amsfonts}
\usepackage{amsmath} 
\usepackage{amssymb} 
\usepackage{mathrsfs}
\usepackage{parskip}

\usepackage{color}

\definecolor{naranja}{rgb}{1,0.5,0}

\def\simlt{\stackrel{<}{{}_\sim}}

\def\TeV{\, {\rm TeV}}

\newcommand{\be}{\begin{eqnarray}}
\newcommand{\ee}{\end{eqnarray}}

\usepackage{amssymb}

\usepackage[figuresright]{rotating}
\usepackage[font=small,labelsep=none]{caption}

\begin{document}

\begin{frontmatter}




\title{SUSY and BSM in the face of LHC-14}


\author{J. Alberto Casas}

\address{Instituto de F\'{\i}sica Te\'orica, IFT-CSIC/UAM, Madrid, Spain}

\begin{abstract}
In this talk I review the motivations for physics beyond the Standard Model at the TeV scale and the prospects for their detection in the second Run of LHC. Then I focus in the supersymmetric case, paying special attention to the formulation and implications of the Natural SUSY scenario.

{\em Contribution to the proceedings of the X SILAFAE, Medellin, Colombia, November 24-28 2014.}
\end{abstract}


\end{frontmatter}


\section{Heritage from LHC-8}\label{}
In the first run of LHC we have seen an historical success, the discovery of the Higgs boson. We all hope that in the second run, at 13-14 TeV, we will also witness another historical success, namely the discovery of physics beyond the Standard Model (BSM), which is actually the second main goal of the LHC project. Certainly, the first LHC run has not produced any serious hint of BSM physics, though there were reasonable expectations in that sense. First of all, the Higgs represents the last sector of the SM and thus the last "terra incognita" where a big surprise (i.e. a major departure from SM expectations) could take place. Second, the main theoretical arguments to expect BSM at LHC rely on the naturalness of the electroweak symmetry breaking, so Higgs physics is a natural arena to find new physics. However, fortunately or not, we have not found any sign of new physics. Higgs properties appear so far in great agreement with SM predictions, as illustrated by Fig.1,which summarizes the CMS results on Higgs observables \cite{CMS:2014ega}. Incidentally, LHC has not only discovered the Higgs boson, but also new forces, the Yukawa interactions (for the top, bottom and tau). This represents another major discovery, again in perfect agreement with the SM. The conclusion is that if there is BSM physics related to the Higgs, it must be hidden within the error bars! This seems wishful thinking. However, although LHC results are very impressive, error bars are still sizable and, besides, very important properties of the Higgs are still to be tested, e.g. its coupling to muons (from $H\rightarrow \mu\mu$, hopefully at the reach of LHC-14) and its self-coupling (difficult to measure at LHC).
\begin{figure}[ht]
\centering 
\includegraphics[width=0.8\linewidth]{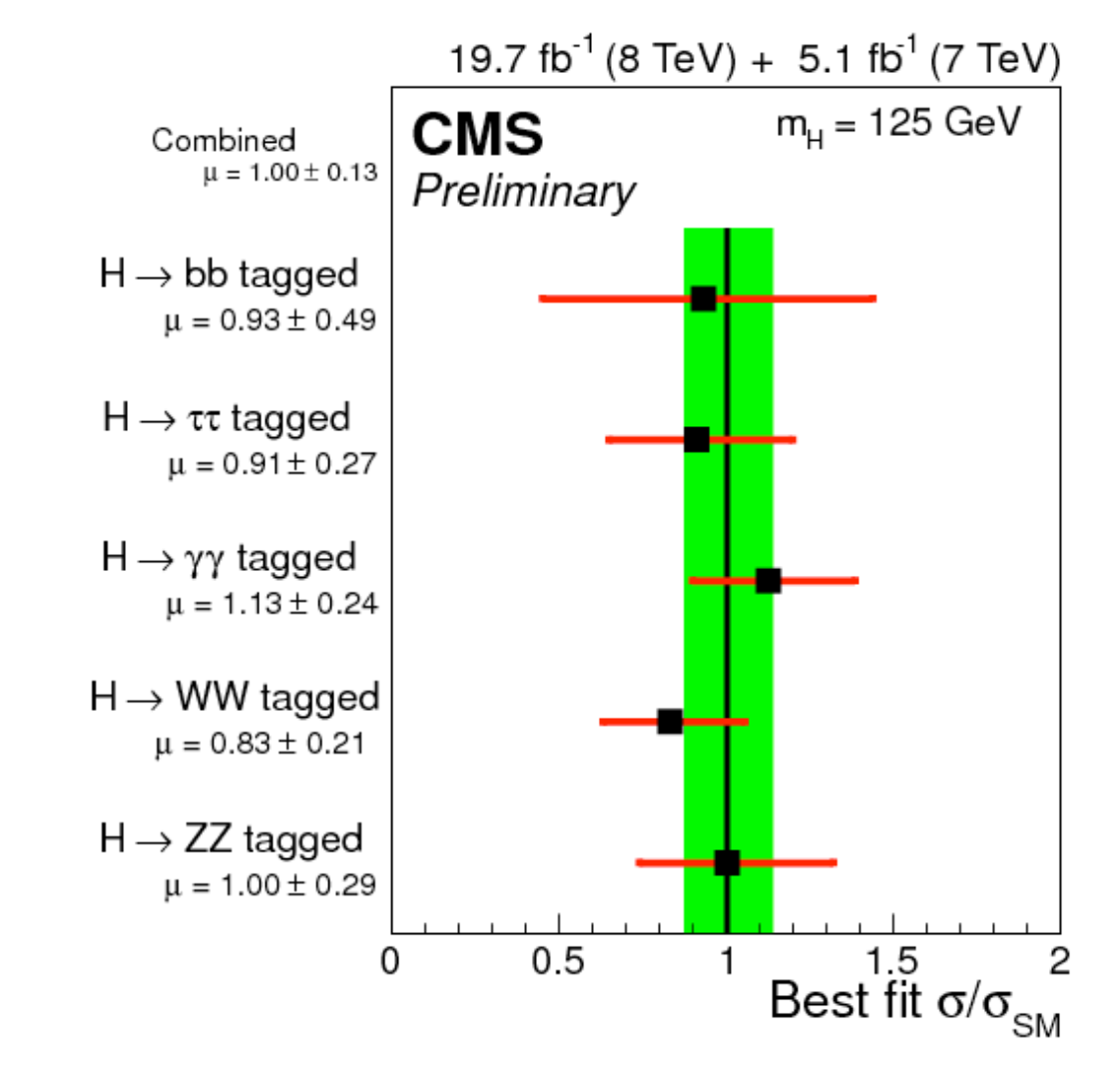}
\caption{}
\label{fig:CMSsummary}
\end{figure}
Apart from Higgs physics, LHC has shown a dramatic agreement with SM predictions in an enormous variety of processes.

\section{The tension between LHC and the naturalness argument}\label{}

The absence of hints of new physics from the Run I of LHC has created in the community the sense that there is a tension between the LHC results and the naturalness arguments behind the (usual interpretation of the) hierarchy problem. How strong is this tension depends on the person you talk to, which reflects the slippery character of the subject. So, let us briefly recall, once more, the old hierarchy-problem argument. You just compute the (one-loop) quadratic contributions to the Higgs-mass parameter, $m^2=m_h^2/2$. Using a cut-off regularization, they read 
\be
\label{HP1}
\delta m^2=\frac{\Lambda^2}{4\pi v^2} (-3m_t^2+\cdots)\ ,
\ee
where $v=\langle H \rangle$ and we have written only the top contribution, which is the most important one due to its large Yukawa coupling. Then, if you demand that this contribution is not much larger than the mass parameter itself, say less than 10 times larger, you get  an upper bound on the cut-off, $\Lambda$,
\be
\label{HP2}
\left|\frac{\delta m^2}{m^2}\right|\leq 10 \ \ \ \ \Rightarrow \ \ \ \ \Lambda \lesssim 1.5 \TeV
\ee
(the subdominant contributions are positive, making the bound slightly weaker). It is impressive that this simple argument is still the main theoretical reason to expect BSM physics at the reach of LHC. Admittedly, this naturalness criterion is quite imprecise and maybe too-naive, or even a misconception. But, given its importance, we may wonder to which extent the LHC results are in tension with it.  In this sense, notice that, certainly, the LHC has explored a lot of physics up to 1.5 TeV, but not all the physics. Since the above naturalness bound comes from the top contribution, it applies to BSM physics associated to the top, i.e. physics able to cancel the top contribution in some way.  But the fact is that, even if there are indeed ``top partners" of any kind with masses at e.g. 1 TeV, they could have easily escaped the Run I of LHC. For instance, if the BSM physics is supersymmetry (SUSY), the top partners are stops. But the present ATLAS and CMS limits allow stops at 700 GeV, or even lighter depending on the topology of the stop decay, see Fig. 2 \cite{ATLAS}.
\begin{figure}[ht]
\centering 
\includegraphics[width=1.0\linewidth]{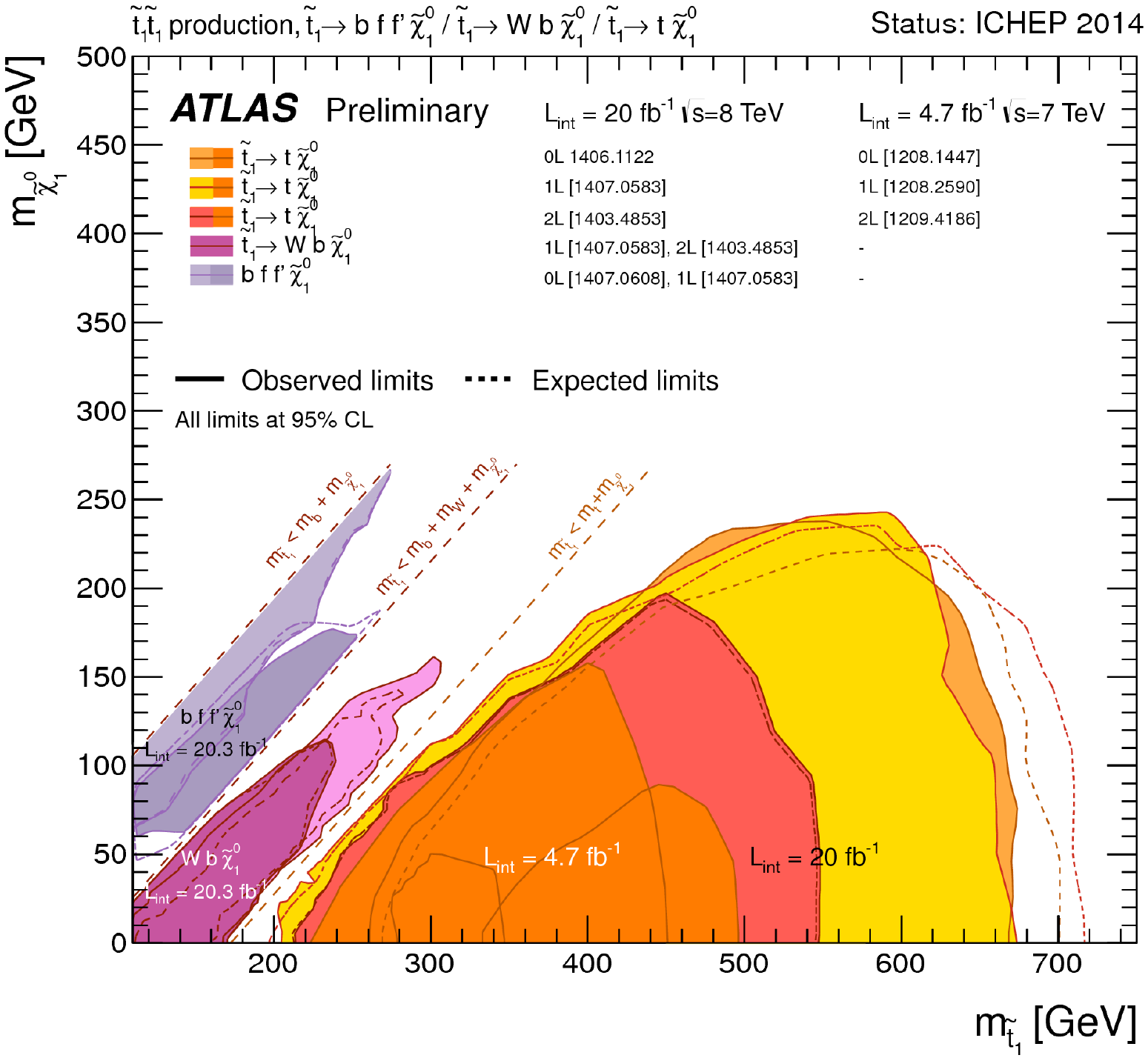}
\caption{}
\label{fig:stops}
\end{figure}
Hence  it is maybe a bit too soon to give up. It could even happen that the naturalness criterion is sound, but the new physics is just above the reach of the Run II of the LHC (hope not). Notice also that the  -rather vague- naturalness bounds become more precise when they are analyzed for concrete BSM scenarios, since in that case we can evaluate the various contributions to the Higgs mass in terms of the initial parameters of the theory, and thus study the actual degree of cancellation which is required to keep the Higgs mass parameter at the correct value (we will see an example of this later). 


So there are two possibilities. The first one is that the naturalness criteria behind the hierarchy problem do {\em not} apply. This can be either because we have misconceptions about the hierarchy problem, or maybe because the electroweak scale is fixed by anthropic reasons within a landscape framework. Furthermore, there have appeared suggestions, like A-gravity \cite{Salvio:2014soa},  that the stability of the electroweak scale does not require BSM physics. In any of these cases LHC (or even more powerful colliders) will probably fail to find any BSM physics. The second possibility is that the naturalness argument applies. In that case we can expect BSM physics at the TeV scale, probably (if we are lucky) at the LHC reach.
Of course we do not know yet which way has been chosen by nature, see Fig. 3. 
\begin{figure}[ht]
\centering 
\includegraphics[width=1.0\linewidth]{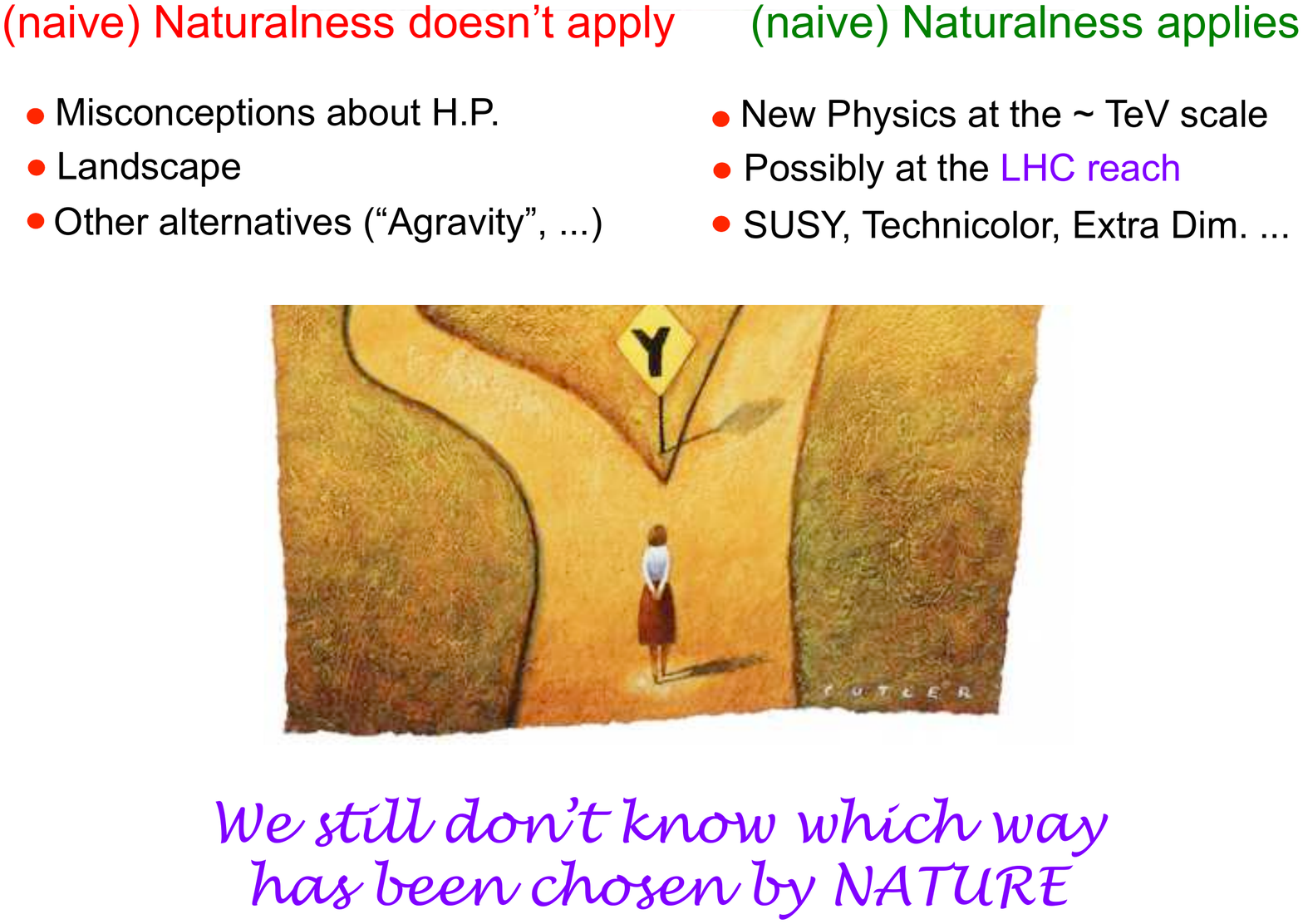}
\caption{}
\label{fig:nature}
\end{figure}
Therefore, at the moment it seems reasonable to explore both possibilities. For the rest of the paper I will assume that the naturalness argument is sound, so we can expect new physics at the TeV scale. Which one?


\section{Which BSM physics?}\label{}

There many models of BSM physics which address the hierarchy problem. Probably the best motivated ones are Supersymmetry, Composite/NGB Higgs models and (warped) extra dimensions. They are attractive because their theoretical foundations are solid and were not developed ad hoc to cure the hierarchy problem; surely, which one you prefer is a matter of taste. In fact, these are not models but frameworks that include many different models inside. 
It is fair to say that, though these frameworks are well motivated, when one goes to the details, they all look (much) uglier. Maybe this is a bad signal for the most conventional BSM frameworks, or maybe we are not ingenious enough to devise more satisfactory models. (Or perhaps our standards of beauty do not match nature's.). 


There are two main strategies to look for new physics: direct searches of new particles and processes, and indirect searches based on the fingerprints of the new physics in the effective (SM-like) theory. The last one has received a lot of attention in the last two years (see talk by C. Grojean in this conference). One considers the most general effective Lagrangian based on the SM, extended with higher-order operators involving SM fields.
Then, one looks for deviations on SM observables which could signal the presence of these extra operators, and thus of new physics. Which strategy, direct or indirect, is more efficient depends on the kind of BSM physics which is actually there (if any). We will see examples in both senses soon. Actually, in many cases the two strategies are complementary, since possible direct (indirect) signals can be cross-checked or falsified by indirect (direct) ones. So the most reasonable attitude is to explore both avenues. Incidentally, the same crossroad arises when decisions are to be made about future accelerator programs (e.g. up-graded LHC for direct searches or ILC for indirect ones).

Though most of this short review is devoted to SUSY, let us say a few words about composite-Higgs models, which also illustrate some of the previous points. Composite-Higgs models are based on the hypothesis that the Higgs boson is a composite state of fermions, glued by new (strong) interactions. Present models of this kind usually incorporate the Higgs boson as the pseudo-Nambu-Goldstone-boson of some spontaneously broken symmetry \cite{Dimopoulos:1981xc}. In this way, the Higgs boson is naturally lighter than other states, which live at the scale where the new interactions are strong. This is similar to the mechanism that makes the pion lighter than the other hadronic states. The appeal of this scheme is precisely the fact that it has been already realized by nature with the ordinary strong interactions. The hierarchy problem is solved because above the TeV scale the Higgs-boson does not appear as an elementary scalar anymore, so there are not large quadratic radiative-corrections to its mass.

However, the scheme has problems. To reproduce the top mass requires couplings of the top to heavy (composite) fermions, which would be the "top partners", with ${\cal O}$(TeV) masses \cite{Kaplan:1991dc}. The Lagrangian must contain terms like
\be
\mathscr{L}_Y\supset \lambda_L \overline u_L \mathcal{Q}_{R} \ +\ \lambda_R \overline{\mathcal{U}}_{L}u_R
\ee
where $\cal{Q} , \cal U$ are the top (composite) partners. This implies that the physical top is in fact a partially composed particle. All this game produces modified couplings of the Higgs to the SM particles. Consequently, the exploration of the Higgs effective field theory (which is a part of the SM-like effective theory) seems the most promising way to find {\em this kind} of new physics. Incidentally, many models conceived in the context of warped extra dimensions are equivalent (through AdS/CFT correspondence) to composite models. So, their phenomenology is also similar.

Before going to SUSY, it is worth mentioning that, beside the hierarchy problem, there are other reasons to believe in the existence of BSM physics, hopefully at the reach of LHC or other experiments. The most important one is dark matter, which is actually the strongest evidence for BSM physics. If dark matter is made of WIMPs (one of the best candidates for such starring role), then their mass should be in the O(10 GeV) - O(1 TeV) range, perhaps at the reach of LHC.  Another important motivation for BSM is flavour physics. We do not know at which scale lies the underlying physics for the mysterious observed pattern of fermion masses and mixing angles. Present bounds on flavour-violating operators in the effective SM-like theory indicate that such scale could be much larger than the energy-reach of LHC (for direct detection). This means that for flavour physics, indirect searches are a more efficient way of discovery. Of course this happens because in the SM flavour-violating processes are severely suppressed and also because there are high-quality experimental limits on them. 
The searches for signals of dark matter and flavour-violating processes are an important part of the LHC program. Certainly, we would be lucky if the physics related to these two issues shows up explicitely at the LHC, but even if they do not, LHC has already shown its power to put useful limits in models of dark matter and flavour, which are complementary to those from other experiments. Those limits will continue to improve in the Run II (see talks by A. Ibarra and Y. Nir in this conference).
Actually, both ATLAS and CMS will keep on the search for other kinds of BSM physics which do not have any special motivation, but could simply be there. Examples of this are the search for $Z'$s, extended Higgs sectors (e.g. with another doublet or a singlet), universal extra-dimensions (at the TeV), etc. It would be a fortunate situation if any of this new physics is found at the LHC, but it could happen, as it has happened in the past in similar situations; recall the "unexpected discoveries" of the muon in 1937 or the dark energy in 1998. Let us also mention that if the present ($\sim 3 \sigma$) discrepancy between theory and experiment for $(g-2)_\mu$ \cite{Dorokhov:2014iva} is eventually confirmed, this would be a clear signal of new physics, most probably around the TeV scale. 

\section{SUSY}\label{}

SUSY has been the paradigmatic scenario of BSM physics, for good theoretical and phenomenological reasons, well-known to everybody. SUSY is a beautiful symmetry, strongly suggested by string theories. It provides an elegant solution to the hierarchy problem thanks to the cancellation of the dangerous quadratically-divergent contributions to the Higgs mass. 
In addition, SUSY presents nice features which was not designed for.  The fact that, so far, the Higgs boson looks fundamental is in agreement with the supersymmetric expectations. Even more important, the Higgs mass, $m_h\simeq 125$ GeV is below the upper bound $m_h\lesssim 135$ GeV, arising in the minimal supersymmetric standard model (MSSM), which the simplest supersymmetric extension of the SM. A Higgs of 180 GeV (perfectly natural in the SM context) would have been devastating for SUSY. 
Also, in the context of SUSY the electroweak (EW) breaking occurs due to radiative effects in a quite natural way, since the square-mass of one of the two Higgs doublets is driven towards negative values along the renormalization group running from the high scale. Furthermore, SUSY models offer a perfect WIMP candidate for dark matter, namely the lightest supersymmetric particle, which is normally the lightest neutralino. Finally, the supersymmetric scenarios show a beautiful unification of the gauge couplings at the unification scale, $M_X\simeq 2\times 10^{16}$ GeV, not far from the Planck scale. Everything is nice, {\em but} there are problems. First, a Higgs at 125 GeV is a bit too heavy for naive supersymmetric expectations. Arranging such a Higgs mass requires, at least within the MSSM, heavy stops and thus some degree of fine-tuning (more details later). Second, we have not seen any signal of SUSY at the Run I of LHC. Of course this fact is shared by all the BSM models. The combination of the two previous facts implies that all the supersymmetric particles must be of order 1 TeV (or larger). This leads to a fine tuning to get the correct value of the EW scale, typically at the $0.1\%-1\%$ level. Again, this fine-tuning trouble is not only for SUSY. It generically occurs for any BSM physics potentially capable of solving the hierarchy problem.

In contrast to what happened for composite-Higgs models, in the case of SUSY direct searches are typically more efficient than indirect ones. The reason is that, due to R-parity, SUSY-induced diagrams in the SM-like effective theory (i.e. containing just SM particles in the external legs) must have a loop, since all the new interaction vertices involve two supersymmetric particles. Thus all the additional higher-order operators generated in the effective theory have a loop suppression-factor. E.g.
\be
\mathscr{L}_{\rm eff}\sim \frac{g^6}{(4\pi)^2} \frac{1}{\Lambda^2}|H|^6 \ +\ \cdots
\ee
This means that, in a very natural way, SUSY is decoupled from the low-energy physics. If SUSY is really there, this would explain why it has not been seen yet in the SM-like observables.

Concerning direct searches, the highest cross sections of SUSY production are normally gluino and/or squark pair-production. By squark we mean here a squark state of the first or second generation since the production of stops and sbottoms is very suppressed. In a typical SUSY process the gluinos ($\tilde g$) and squarks ($\tilde q$) decay along cascades with diverse topology. In models with some kind of R-parity (desirable though not mandatory to avoid proton decay and other baryon/lepton number violating processes), each cascade always produces one lightest supersymmetric particle (LSP), typically a neutralino $\chi_1^0$, among the final states. In addition, one or more jets, with or without leptons, are created in each cascade. Therefore the most direct search for SUSY is to look for events with

\begin{itemize}
\item jets with high $p_T$
\item $E_T^{\rm miss}$
\item 0-N leptons
\end{itemize}

\noindent Normally, multijet events with large $E_T^{\rm miss}$ and 0-leptons are the most efficient channel, but channels with leptons may play an important (sometimes capital) role, especially for particularly elusive types of SUSY spectrum.

It is not straightforward to translate LHC data 
into concrete limits on SUSY (MSSM) parameters. Let us recall that the MSSM has $\sim 100$ independent parameters, mainly soft terms related to the unknown mechanism of SUSY breaking and its transmission to the observable sector, $\{m_{ij}^2,M_a,A_{ij},B,\mu\}$.
Here $m$, $M$ and $A$ are scalar masses, gaugino masses and trilinear scalar couplings; $i,j$ and $a$ are family and gauge group indices respectively; $B$ is the bilinear scalar coupling and $\mu$ is the usual Higgs mass term in the superpotential. Requiring no flavour or CP violation in the first and second generations reduces the number of parameters to $\sim 20$ (plus those already present in the SM), still a huge number. It is certainly cumbersome to translate the LHC data into constraints on such complex parameter-space. A usual strategy is to present the LHC data as constraints in a simpler, but well motivated, version of the MSSM (more on this below). An alternative procedure, which is gaining relevance, 
 is the use of so-called ``simplified models". A simplified model is defined by an effective Lagrangian describing the interactions of a small number of new particles,.
 They can thus be defined by a small number of masses and cross-secctions. The latter parameters are directly related to collider-physics observables. The idea is to mimic the collider signatures of a particular physical scenario (e.g. some concrete version of the MSSM) with a dominant simplified model (or a reduced set of them) in each region of the parameter space. This makes more efficient the exploration of such complex models. E.g. squark or gluino decays $\tilde q \rightarrow q \chi_1^0$, $\tilde g \rightarrow q \bar q \chi_1^0$ are dominant if the other relevant super-particles are heavier. In a simplified model the masses of the latter can be just sent to infinity. Of course, additional complexity can be built-in. The strategy is also efficient for non-supersymmetric models of new physics. 

Though simplified models are very useful, one has to be careful in the interpretation of the experimental results when they are given in terms of them. One has to read the small letters of the plots, otherwise one could be led to wrong conclusions. As an example, consider the present and future bounds on electroweakinos (charginos and neutralinos) from the tri-lepton signal. The exclusion limits from ATLAS on the mass of the lightest chargino are given in Fig.~4 \cite{ATLAS}, which shows the present limits (brown region) and the future ones. 
\begin{figure}[ht]
\centering 
\includegraphics[width=1.0\linewidth]{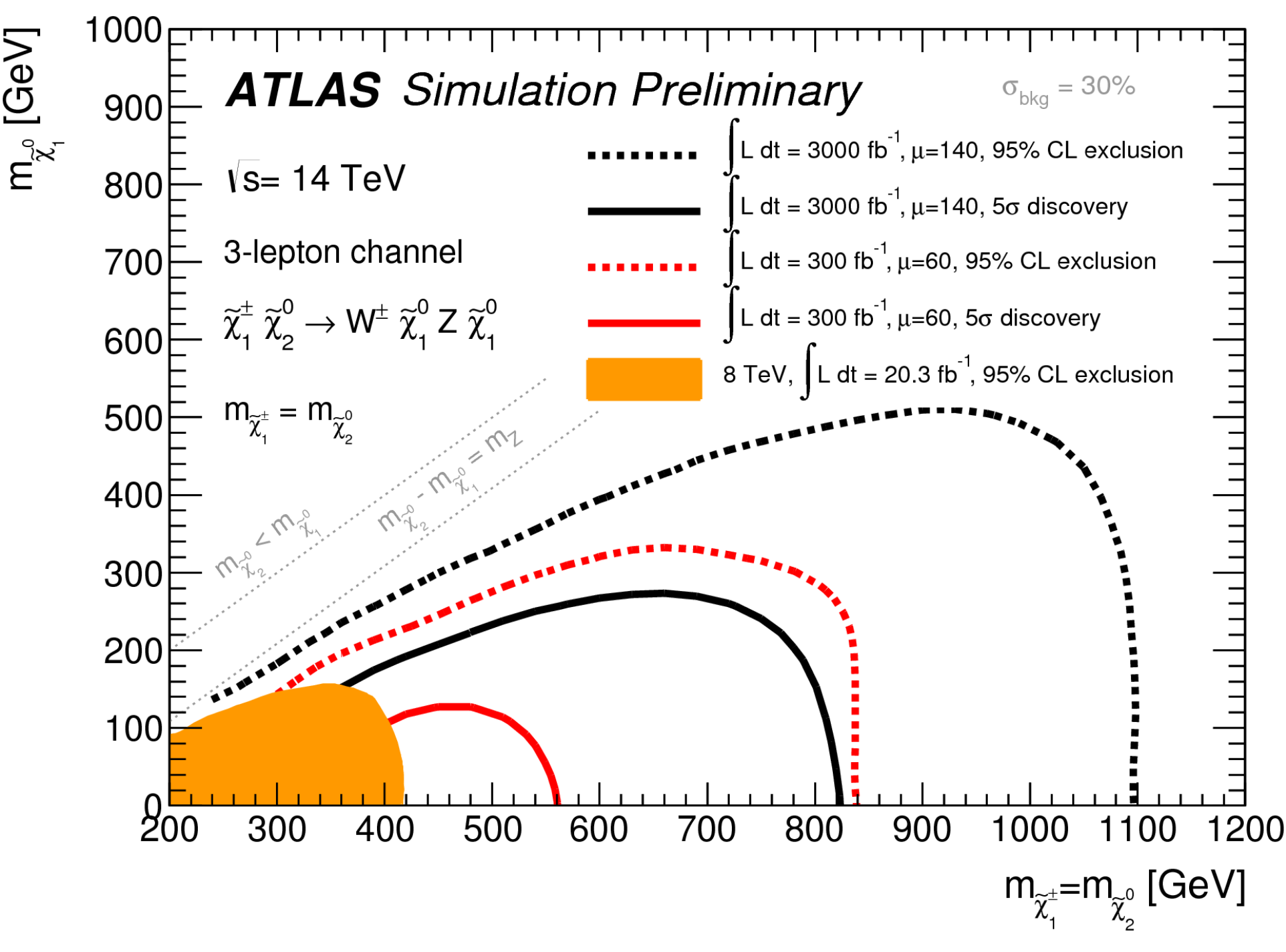}
\caption{}
\label{fig:electroweakinos}
\end{figure}
The plot is very impressive, since in the future we will have limits on charginos at $\sim 1$ TeV. However, the plot has been obtained under certain simplifying assumptions. Namely, it has been assumed that the lightest chargino is degenerate with the second lightest neutralino. Moreover, the decays of these particles
are supposed to occur through a $W-$ and a $Z-$boson respectively. These are common instances when the lightest neutralino is mostly bino, and the lightest chargino and the second lightest neutralino are mostly winos. However, if e.g. the second lightest neutralino is mostly Higgsino, it typically decays through a Higgs, in which case it is much more difficult to identify the chargino/neutralino production and the bounds become very weak.

Let us now consider the other strategy to present the LHC constraints in SUSY, namely to translate the LHC data into limits on a simpler, well motivated, version of the MSSM. In this sense, the most heavily used model is the constrained MSSM (CMSSM).  Then the previous (100 or 20) parameters are reduced to $\{m,M,A,B,\mu\}$, i.e. the universal scalar mass, gaugino mass and trilinear scalar coupling; plus the $B$ and $\mu$ parameters. All quantities are to be understood at the high scale $M_X$. Using the EW breaking conditions, coming from the minimization of the Higgs potential, one can eliminate $\mu$ (except for the sign) and trade $B$ by $\tan\beta=\langle H_u\rangle/\langle H_d\rangle$ (the ratio of the expectation values of the two Higgs doublets). So the usual set of CMSSM parameters is  
\be
\left\{m,M,A,B,\tan\beta, {\rm sign}\mu\right\}
\nonumber
\ee
Due to the remarkable growth of the gluino and squark masses along the RG running, the typical CMSSM spectrum is
\be
&M_{\tilde g}& \sim m_{\tilde q} > m_{\tilde l}
\nonumber\\
&M_{\tilde g}& > M_{\chi^\pm} \gtrsim M_{\chi^0_1}
\nonumber\\
&{\chi^0_1}&\ \equiv\ {\rm LSP}
\nonumber
\ee
where $\chi^\pm$ and $\chi^0_1$ denote the lightest chargino and neutralino states (the squark can be much heavier than the gluino if $m\gg M$). The values of $A$ and $\tan\beta$ are not very important for the multijet signal, as they play  almost no role in the gluino and squark production. Because of that, the impact of the LHC results on the CMSSM are usually presented as exclusion limits in the $m-M$ plane (or in the $m_{\tilde q}-M_{\tilde g}$ plane) for $A$ and $\tan\beta$ fixed.
Fig.~5 shows the last ATLAS analysis  \cite{ATLAS}. The results for CMS are similar\cite{CMS}.

\begin{figure}[ht]
\centering 
\includegraphics[width=1.0\linewidth]{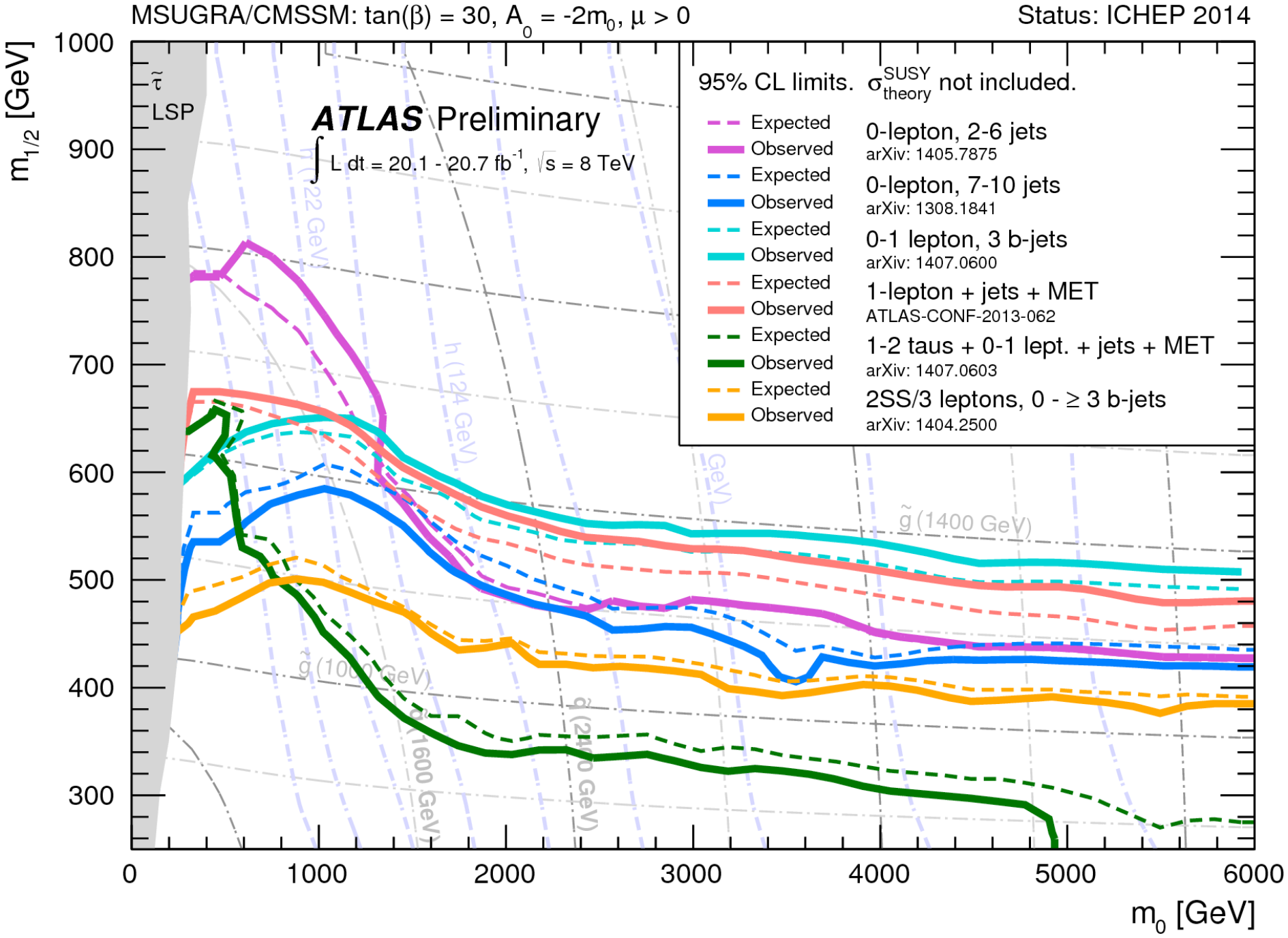}
\caption{}
\label{fig:CMSSM}
\end{figure}
Roughly speaking, the results imply $m_{\tilde q}\ \gtrsim\ 1800\ {\rm GeV}$, $M_{\tilde g}\ \gtrsim\ 1400\ {\rm GeV}$. 

On the other hand, the experimental Higss mass, $m_h\simeq 125$ GeV has a deep impact in the MSSM spectrum. 
As  is well known, the tree-level Higgs mass in the MSSM is given by $(m_h^2)_{\rm tree-level}=M_Z^2 \cos^22\beta$. Since its maximum value (achieved for $\tan\beta\gtrsim 5$) is the mass of Z-boson, radiative corrections are needed in order to reconcile it with the experimental value. A simplified expression  of such corrections \cite{ Haber:1996fp, Casas:1994us,Carena:1995bx}, useful for the sake of the discussion, is
\begin{equation}\label{eq:der2}
\delta m_{h}^{2}=\frac{3 G_{F}}{\sqrt 2 \pi^{2}}m_{t}^{4}\left(\log\left(\frac{\overline m_{\tilde t}^{2}}{m_{t}^{2}}\right)+\frac{X_{t}^{2}}{\overline m_{\tilde t}^{2}}\left(1-\frac{X_{t}^{2}}{12\overline m_{\tilde t}^{2}}\right)\right)\ +\ \cdots\ ,
\end{equation}
with $\overline m_{\tilde t}$ the average stop mass and $X_{t}=A_{t}-\mu \cot\beta$. The $X_t$-contribution arises from the threshold corrections to the quartic coupling at the stop scale. This correction is maximized for $X_{t}=\pm\sqrt 6 \overline m_{\tilde t}$ ($X_{t}\simeq  \pm2 \overline m_{\tilde t}$ when higher orders are included). Then, unless the trilinear coupling is close to this maximizing value, the requirement of the correct Higgs mass demands heavy stops, $m_{\tilde t}\gtrsim 1$ TeV). Notice that this argument is valid not only for the CMSSM but for any MSSM. 

The experimental bounds on the MSSM spectrum coming from the analysis of the CMSSM, namely $m_{\tilde q}\ \gtrsim\ 1800\ {\rm GeV}$, $M_{\tilde g}\ \gtrsim\ 1400\ {\rm GeV}$ and $m_{\tilde t}\gtrsim 1$ TeV (unless $A_t$ is close to its maximizing value) are in fact quite similar for other supersymmetric models, so the fine-tuning problems that give rise (discussed in the next section), are indeed quite general problems of SUSY, at least of the MSSM. Certainly, there are possible exceptions if SUSY lives in special corners of the parameter space,
e.g. if the SUSY spectrum is ÒcompressedÓ \cite{martin2}, so that visible particles in the events have small $p^T$. Such situation would fool the LHC to some extent, so a much lighter  supersymmetric spectrum (and therefore less problematic for fine-tuning issues) could occur. It is possible, but it sounds artificial, a ÒtrickÓ ad hoc to save low-energy SUSY.

\section{The trouble with the fine-tuning}


Large supersymmetric masses (especially gluino and stop masses) generally imply a rather severe fine-tuning to obtain the correct EW scale. Let us briefly recall how this comes about.

In the MSSM, the vacuum expectation value of the Higgs, $v^2/2 = |\langle H_u\rangle|^2 +|\langle H_d\rangle|^2$, is given, at tree-level, by the minimization relation
\be
\label{mu}
-\frac{1}{8}(g^2+g'^2)v^2 = -\frac{M_Z^2}{2}=\mu^2-\frac{m_{H_d}^2 - m_{H_u}^2\tan^2\beta}{\tan^2\beta-1} \ .
\ee
In the limit of large $\tan\beta$ (which is the relevant one to reproduce the Higgs mass without the need of gigantic stop masses \cite{Cabrera:2011bi}) relation (\ref{mu}) gets simplified
\be
\label{min}
-\frac{1}{8}(g^2+g'^2)v^2 = -\frac{M_Z^2}{2}=\mu^2+m_{H_u}^2\ .
\ee
The absolute value of $m_{H_u}^2$ is typically much larger than $M_Z^2$, so a non-trivial cancellation between $m_{H_u}^2$ and $\mu^2$ is required. This is the fine-tuning associated to the electroweak breaking. (The good news is that generically the sign of  $m_{H_u}^2$ is negative, and thus opposite to that of $\mu^2$.)
%
Note that one cannot just take $m_{H_u}^2$ small while  $M_{\tilde g},\ m_{\tilde t}$, etc. are large, since $m_{H_u}^2$ receives large (and negative) radiative contributions proportional to the latter along the RG running from $M_X$ to $M_{EW}$. In other words, large $M_{\tilde g},\ m_{\tilde t}$,... generally imply large $m_{H_u}$ (in absolute value). E.g. for moderately large $\tan\beta$ 
\be
\label{mhunum}
\hspace{-0.5cm}
m_{H_u}^2&=& -1.6M_3^2 +0.63m_{H_u}^2 -0.37 m_{Q_3}^2 -0.29m_{U_3}^2
\nonumber\\
&+ &  0.28A_tM_3+ 0.2M_2^2+\cdots \; .
%
\ee
Thus the EW breaking is fine-tuned unless the typical soft masses are  ${\cal O}(1 {\rm TeV})$ or less. Incidentally, this fine-tuning is stronger than indicated by the naive hierarchy-problem argument, see eqs.~(\ref{HP1}, \ref{HP2}) and the discussion below. The reason is the large enhancement of the contributions due to long running from $M_X$ to low-energy. If SUSY breaking is transmitted below $M_X$ the fine-tuning would be alleviated. On the other hand, it is well-known that the radiative corrections to the Higgs potential reduce the fine-tuning \cite{de Carlos:1993yy}. This effect can be honestly included taking into account that the effective quartic coupling of the SM-like Higgs runs from its initial value at the SUSY threshold, $\lambda(Q_{\rm threshold})=\frac{1}{8}(g^2+g'^2)$, until its final value at the electroweak scale, $\lambda(Q_{EW})$.
The effect of this running is equivalent to include the radiative contributions to the Higgs quartic coupling in the effective potential, which increase the tree-level Higgs mass, $(m_h^2)_{\rm tree-level}=2\lambda(Q_{\rm threshold})v^2=M_Z^2$, up to the experimental one, $m_h^2=2\lambda(Q_{EW})v^2$. Therefore, replacing $\lambda_{\rm tree-level}$ by the radiatively-corrected quartic coupling is equivalent to replace $M_Z^2 \rightarrow m_h^2$ in eq.(\ref{min}) above \cite{Casas:2014eca}, i.e.
\be
\label{minh}
-\frac{m_h^2}{2}=\mu^2+m_{H_u}^2\ .
\ee
Since $m_h>M_Z$, the fine-tuning is somewhat reduced. For quantitative studies of the fine-tuning, one needs a criterion to parametrize its severeness. A standard parametrization of the degree of fine-tuning \cite{Ellis:1986yg, Barbieri:1987fn},  reads
\be
\label{BG}
\frac{\partial m_h^2}{\partial \theta_i} = \Delta_{\theta_i}\frac{ m_h^2}{\theta_i}\ , \ \ \ \ \ \Delta\equiv {\rm Max}\ \left|\Delta_{\theta_i}\right|\ ,
\ee
where $\theta_i$ are the independent parameters that define the model under consideration and $\Delta_{\theta_i}$ are the associated fine-tuning parameters. In SUSY models $\theta_{i}$ are usually taken as the initial (high-energy) values of the soft terms and the $\mu$ parameter. Nevertheless, for specific scenarios of SUSY breaking and transmission to the observable sector, the initial parameters might be particular theoretical parameters that define the scenario and hence determine the soft terms, e.g. a Goldstino angle in scenarios of moduli-dominated SUSY breaking. Roughly speaking $\Delta^{-1}$ is the (p-value) probability that the Higgs VEV is equal or smaller than the experimental value. For a discussion of the statistical meaning of $\Delta_{\theta_i}$ see refs.~\cite{Ciafaloni:1996zh, Cabrera:2008tj, Casas:2014eca}. 

When one applies this fine-tuning criterion to the CMSSM, using the initial (high-energy) values of $\{m,M,A,B,\mu\}$ as the independent parameters, one discovers that the EW breaking is fine-tuned at the 1\% level. Notice that we should not ignore the fine-tuning problem, since the main reason to consider low-energy SUSY was precisely to avoid the hierarchy problem, which is in fact the fine-tuning problem associated to the EW breaking in the SM. (Of course the latter fine-tuning is orders of magnitude more severe than the supersymmetric one.). 

To which extent the fine-tuning problems of the CMSSM remain in general MSSMs? Are there natural way-outs to this situation?
To address these questions let us recall first the original motivations for the CMSSM. These were 1) Minimal Flavour and CP violation, 2) Simplicity, 3) The fact that it arises in some theoretically motivated scenarios (like minimal SUGRA or Dilaton-dominated SUSY breaking). From these motivations only the first one is robust; but, in fact, the experimental constraints do not require fully universal soft terms. E.g. the third generation of squarks and sleptons could have very different masses. The degeneracy of gaugino masses at $M_X$ is not experimentally justified either. Therefore, going beyond the CMSSM is very plausible. On the other hand, since fine-tuning seems to be the main problem with SUSY (actually the only one!), a reasonable guide to explore more general supersymmetric models is to look for scenarios as little fine-tuned as possible. And this is precisely the definition of Natural SUSY.

\section{Natural SUSY}

Naturalness arguments have been used since long ago \cite{Barbieri:1987fn}  to constrain from above supersymmetric 
masses.
Already in the LHC era, they were re-visited in ref.~\cite{Papucci:2011wy} to formulate the so-called Natural SUSY scenario. This has become very popular in the last times, as a framework that gives theoretical support to searches for light stops and other particles at the LHC, a hot subject from the theoretical and the experimental points of view. In that paper the authors evaluate the most important contributions to $m_{H_u}^2$, in a certain approximation. E.g. the stop contribution was taken as  
\begin{equation}\label{eq:der1}
\delta m_{H_{u}}^{2}|_{\rm stop}=-\frac{3}{8\pi^{2}}y_{t}^{2}\left(m_{Q_{3}}^{2}+ m_{U_{3}}^{2}+|A_{t}|^{2}\right)\log\left(\frac{\Lambda}{\TeV}\right),
\end{equation}
where $\Lambda$ denotes the scale of the transmission of SUSY breaking to the observable sector and  the 1-loop leading-log (LL) approximation was used to integrate the renormalization-group equation (RGE). Then, the above soft parameters $m_{Q_{3}}^{2}$, $m_{U_{3}}^{2}$ and $A_{t}$ are to be understood at low-energy, and thus  they control the stop spectrum. This sets an upper bound on the stop masses, according to which stops should be quite light ($\lesssim 1$ TeV)

However, the previous one-loop LL approximation is too simplistic in two different aspects.
First, it is not accurate enough since the top Yukawa-coupling, $y_t$, and the strong coupling, $\alpha_s$, are large and vary a lot along the RG running. 
Second, the physical squark, gluino and electroweakino masses are not initial parameters, but rather a low-energy consequence of the initial parameters at the high-energy scale. This means that one should evaluate the cancellations required among those initial parameters in order to get the correct electroweak scale.
This  entails two complications. First, there is not one-to-one correspondence between the initial parameters and the physical quantities, since the former get mixed along their coupled RGEs. And second, the results depend (sometimes critically) on which parameters one considers as the initial ones.

A most relevant analytic study concerning this issue is the well-known work by Feng et al. \cite{Feng:1999mn}, where they studied the focus 
point  region of the CMSSM . In the generic MSSM, the (1-loop) RG evolution of a shift in the initial values of $m_{H_u}^2, m_{U_3}^2, m_{Q_3}^2$ reads 
\begin{equation}
\frac{d}{dt} \left[ \begin{array}{c} \delta m_{H_u}^2
\\ \delta m_{U_3}^2 \\ \delta m_{Q_3}^2 \end{array} \right]
= \frac{y_t^2}{8\pi^2} \left[
\begin{array}{ccc}
3 & 3 & 3 \\
2 & 2 & 2 \\
1 & 1 & 1 \end{array} \right]
\left[ \begin{array}{c} \delta m_{H_u}^2
\\ \delta m_{U_3}^2 \\ \delta m_{Q_3}^2 \end{array} \right] \ , 
\end{equation}
where $t\equiv\ln Q$, with $Q$ the  renormalization-scale, and $y_t$ is the top Yukawa coupling.
Hence, starting with the CMSSM universal condition at $M_X$: $m_{H_u}^2=m_{U_3}^2=m_{Q_3}^2=m_0^2$, one finds
\begin {equation}
\delta m_{H_u}^2  = \frac{\delta m_0^2}{2} \left\{3\  
{\rm exp} \left[ \int_0^t \frac{6y_t^2}{8\pi^2} dt' \right]
- 1  \right\}.
\label{focus}
\end{equation}
Provided $\tan\beta$ is large enough, ${\rm exp}\! \left[ {6\over8\pi^2}
\int_0^t y_t^2 dt' \right] \simeq 1/3$ for the integration between $M_X$ and the electroweak scale,  so the value of $m_{H_u}^2$ depends very little (in the CMSSM) on the initial scalar mass, $m_0$. However, the average stop mass is given by 
\begin{eqnarray}
\label{mstops}
\overline m^2_{\tilde t} \simeq 2.97 M_3^2 + 0.50  m_0^2 + \cdots  \ ,
\end{eqnarray}
where $M_3$ is the gluino mass at $M_X$. Therefore, if the stops are heavy {\em because} $m_0$ is large, this does {\em not} imply fine-tuning. This is a clear counter-example to the need of having light stops to ensure naturalness.

From the previous discussion it turns out that the most rigorous way to analyze the fine-tuning is to determine the full dependence of the electroweak scale (and other potentially fine-tuned quantities) on the initial parameters, and then derive the regions of constant fine-tuning in the parameter space. 
These regions can be (non-trivially) translated into  constant fine-tuning regions in the space of possible physical spectra.
This goal is enormously simplified if one determines in the first place the analytical dependence of low-energy quantities on the high-energy initial parameters.
Fortunately, this can be straightforwardly done, since the dimensional and analytical consistency dictates the form of the dependence. E.g. the low-energy (LE) values of $m_{H_u}^2$ and $\mu$ read
%
%
\begin{eqnarray}
\label{mHu_gen_fit}
\hspace{-0.7cm}m_{H_u}^2(LE)&=&
c_{M_3^2}M_3^2 +c_{M_2^2}M_2^2 +c_{M_1^2}M_1^2 + c_{A_t^2}A_t^2
\nonumber\\
&+&c_{A_tM_3}A_tM_3 +c_{M_3M_2}M_3M_2+ c_{m_{H_u}^2}m_{H_u}^2
\nonumber\\
&+& c_{m_{Q_3}^2} m_{Q_3}^2 +c_{m_{U_3}^2}m_{U_3}^2+\cdots
\\
\mu(LE)&=&c_\mu \mu \ ,
\label{mu_gen_fit}
\end{eqnarray}
where all the quantities in the r.h.s. are understood at the high-energy (HE) scale. The values of $c_{M_3^2}, c_{M_2^2},...$ are obtained by fitting the result of the numerical integration of the RGEs to eqs.(\ref{mHu_gen_fit}, \ref{mu_gen_fit}), see ref.~\cite{Casas:2014eca} for the most recent evaluation (some of the $c$'s can be read from eq.(\ref{mhunum})).

A common practice is to consider the (HE) soft terms and the $\mu-$term as the independent parameters, say 
\begin{eqnarray}
\label{Thetas}
\Theta_\alpha=\left\{\mu, M_3,M_2,M_1,A_t, m_{H_u}^2, m_{H_d}^2, m_{U_3}^2, m_{Q_3}^2, \cdots\right\}, 
\nonumber
\end{eqnarray}
which is equivalent to the so-called ``Unconstrained MSSM". Then one easily computes $\Delta_{\Theta_\alpha}$
\begin{eqnarray}
\label{BGTheta}
\Delta_{\Theta_\alpha}=\frac{\Theta_\alpha}{m_h^2}\frac{\partial m_h^2}{\partial \Theta_\alpha} = -2\frac{\Theta_\alpha}{m_h^2}\frac{\partial m_{H_u}^2}{\partial \Theta_\alpha}  \ ,
\end{eqnarray}
which is trivially evaluated using eq.(\ref{mHu_gen_fit}).
The last expression is not valid for the $\mu-$parameter, for which one simply has  $\Delta_{\mu} =-4 c_\mu {\mu^2}/{m_h^2}$.

Note that for any other theoretical scenario, the $\Delta$s associated with the 
genuine initial parameters, say $\theta_i$, can be inmediately written in terms of $\Delta_{\Theta_\alpha}$ 
using the chain rule
\begin{eqnarray}
\hspace{-0.5cm}
\Delta_{\theta_i}\equiv\frac{\partial\ln m_h^2}{\partial \ln\theta_i}
=\sum_\alpha\Delta_{\Theta_\alpha} \frac{\partial \ln \Theta_\alpha}   {\partial \ln\theta_i}
=
\frac{\theta_i}{m_h^2}
\sum_\alpha\frac{\partial m_h^2}{\partial \Theta_\alpha} 
 \frac{\partial\Theta_\alpha}   {\partial \theta_i}\ .
 \nonumber
\end{eqnarray}
Finally, in order to obtain fine-tuning bounds on the parameters of the model we demand $\left|\Delta_{\theta_i}\right| \simlt \Delta^{\rm max}$, where 
$\Delta^{\rm max}$ is the maximum amount of fine-tuning one is willing to accept. E.g. $\Delta^{\rm max}=100$
represents a fine-tuning of $\sim 1\%$. Hence, for the unconstrained MSSM we simply demand
\begin{eqnarray}
\label{FTbound}
\left|\Delta_{\Theta_\alpha}\right| \simlt \Delta^{\rm max}\ ,
\end{eqnarray}
where $\Delta_{\Theta_\alpha}$ are given by eq.(\ref{BGTheta}).
Now, for the parameters that appear just once in  eqs.(\ref{mHu_gen_fit}, \ref{mu_gen_fit}) the corresponding naturalness bound (\ref{FTbound}) is trivial and has the form of an upper limit on the size of the parameter. For dimensional reasons this is exactly the case for mass-dimension-two parameters, e.g. 
\begin{eqnarray}
\label{BGmQ3}
\left|\Delta_{m_{Q_3}^2}\right|=  \left|-2\frac{m_{Q_3}^2}{m_h^2}\  c_{m_{Q_3}^2}\right|\simlt \Delta^{\rm max} \  .
\end{eqnarray}
%
%
This translates into upper bounds on the high-energy soft masses. The bounds on the physical masses require further work (see below).

On the other hand, for dimension-one parameters  (except $\mu$) the naturalness bounds (\ref{FTbound}) appear mixed. In particular, this is the case for the bounds associated to $M_3, M_2, A_t$. From eqs.(\ref{BGTheta}) and (\ref{mHu_gen_fit})
\begin{eqnarray}
\label{BGM3}
\hspace{-0.5cm}
\left|\Delta_{M_3}\right|=\frac{1}{m_h^2}\left|4 c_{M_3^2}M_3^2 +2c_{A_tM_3}A_tM_3+2c_{M_3M_2}M_3M_2\right| 
\nonumber
\end{eqnarray}
\begin{eqnarray}
\hspace{-0.5cm}
\left|\Delta_{M_2}\right|=\frac{1}{m_h^2}\left|4 c_{M_2^2}M_2^2 +2c_{A_tM_2}A_tM_2+2c_{M_3M_2}M_3M_2\right| 
\label{BGM2}
\nonumber
\end{eqnarray}
\begin{eqnarray}
\hspace{-0.5cm}
\left|\Delta_{A_t}\right|=\frac{1}{m_h^2}\left|4 c_{A_t^2}A_t^2 +2c_{A_tM_3}A_tM_3+2c_{A_tM_2}A_tM_2\right| 
\label{BGAt}
\nonumber
\end{eqnarray}
Other parameters, like $M_1, A_b$, get also mixed with them in the bounds, but their coefficients are much smaller.

The next step is to translate the bounds in the initial parameters into limits on the physical supersymmetric spectrum. Therefore, one has to go back from the high-energy scale to low-energy one, using the RG equations. Once more, this can be immediately done using appropriate analytical expressions \cite{Casas:2014eca}. This requires extra work since  there is not a one-to-one correspondence between the physical masses, and the soft-parameters and $\mu-$term at high-energy (for further details see ref.~\cite{Casas:2014eca}). A summary of the most relevant bounds is given in table 1.
\begin{table}[hbt]
 \centering
 {\small
\begin{tabular}{| l | c | c | c |}
\hline
~&$M_{\rm HE}=2\times10^{16}$ &$M_{\rm HE}=10^{10}$& $M_{\rm HE}=10^{4}$ \\
 \hline
$M_{\tilde g}^{\rm max}$ &1 440 & 1 890 &  5 860 \\
$M_{\tilde W}^{\rm max}$ & 1 303 &1 550 &  3 435\\
$M_{\tilde B}^{\rm max}$ & 3 368 & 4 237 &  10 565 \\
$M_{\tilde H}^{\rm max}$ & 626 & 610 &  620 \\
$\overline m_{\tilde t}^{\rm max}$ & 1 650 & 1 973 &  4 140 \\
$ m_{H^0}^{\rm max}$ & 7 252 & 14 510 &  9 900 \\
\hline
\end{tabular}
}
\caption{}
\label{tab:PhysMax}
\end{table}
All the bounds have been obtained by setting $\Delta^{\rm max}=100$, they simply scale as $\sqrt{\Delta^{\rm max}/100}$.
Taking into account the present and future LHC limits, the upper bound on the gluino mass is typically the most stringent one, being at the reach of the LHC (for $\Delta^{\rm max}=100$), unless the high-energy scale is rather low. On the other hand, the gluino bound is  the most sensitive one to the value of $M_{\rm HE}$, since it is a two-loop effect. For $M_{\rm HE}\simeq10^7$~GeV, it is as already beyond the future LHC limit ($\sim 2.5$~TeV)  and it increases rapidly as $M_{\rm HE}$ approaches the electroweak scale. 
The upper bounds on stops are not as stringent as the gluino one unless $M_{\rm HE}$ is pretty close to the electroweak scale, in which case none of them is relevant. In general, it is not justified to say that Natural SUSY prefers light stops, close to the LHC limits. Actually, for $\Delta^{\rm max}=100$ the upper bounds on stops are beyond the LHC reach. Taking lighter stops does not really improve the fine-tuning since there are other contributions to it which are dominant, in particular the gluino one. On the oher hand, if the scalar masses are universal and the HE scale is $M_X$, the previously-discussed focus-point regime makes $m_{H_u}^2$ quite insensitive to the value of the stop mass. Then the upper bound is much higher.

In addition to the EW fine-tuning, there are other potential fine-tunings in the MSSM, namely the tuning of the trilinear coupling to its maximizing value in order to get $m_h=m_h^{\rm exp}$ when stops are too light, and the tuning to get a large $\tan\beta$ (discussed elsewhere \cite{Casas:2014eca}). The first one occurs when the radiative contribution of the stops to $m_h$ is not large enough. Then $A_t$ must be close to $A_t^{\rm max}$ with high precision in order to raise $m_h$ until the experimental value, which entails an additional tuning. So, ironically, too-light stops are disfavoured
from naturalness reasons!
On the other hand, even if there is no fine-tuning to get the experimental Higgs mass, the requirement $m_h=m_h^{\rm exp}$ implies a balance between the radiative and the threshold contributions to $m_h^2$ (see eq.(\ref{eq:der2})), which in turn implies a correlation between the initial parameters, especially $M_3$ (the main responsible for the size of the stop masses, and thus for the radiative contribution) and $A_t$ (main responsible for the threshold one). This correlation has non-trivial consequences for the electroweak fine-tuning, especially taking into accout that the fine-tuning conditions for these parameters are mixed, as discussed above. For more details, see \cite{Casas:2014eca}. 

It is worth mentioning that, concerning the electroweak fine-tuning of the MSSM (i.e. the one required to get the correct electroweak scale), the most robust result is by far that Higgsinos should be rather light, certainly below 700~GeV for $\Delta<100$, i.e. to avoid a fine-tuning stronger than 1\% (all the bounds on masses scale as $\sqrt{\Delta^{\rm max}}$). This result is enormously stable against changes in the HE scale since the $\mu-$parameter (which controls the Higgsino masses) runs very little from HE to LE. The only way it could be substantially relaxed would be that the $\mu-$parameter were theoretically related to the soft masses in such a way that there occurred a cancellation at LE between $\mu^2$ and $m_{H_u}^2$ (see eq.(\ref{minh})). This is difficult to conceive and, certainly, it is not realized in the known theoretical SUSY frameworks. 

On the other hand, the most stringent naturalness upper bound, from the phenomenological point of view, is (as mentioned) the one on the gluino mass. If $M_{\rm HE} \simeq M_X$ one gets $M_{\tilde g}\simlt 1.5$~TeV for $\Delta^{\rm max}=100$, i.e. just around the corner at the LHC. In other words, the gluino mass typically sets the level of the electroweak fine-tuning of the MSSM, which at present is  ${\cal O}(1\%)$.
However, this limit is not as robust as the one on Higgsinos. First, it presents a strong dependence on the HE-scale (due to the two-loop dependence of the electroweak scale on the gluino mass). Actually, for $M_{\rm HE}\simlt10^7$~GeV and $\Delta^{\rm max}=100$ the upper bound on $M_{\tilde g}$ (about 2.7~TeV) goes beyond the present LHC reach. In addition, it could be relaxed if the initial soft parameters (e.g. the gaugino masses) are theoretically related in a favorable way. 

For completness, let us also recall that light stop masses are {\em not} really a generic requirement of Natural SUSY. Actually, stops could be well beyond the LHC limits without driving the electroweak fine-tuning of the MSSM beyond 1\%. Even more, in some scenarios, like universal scalar masses with $M_{\rm HE}=M_X$, stops above 1.5~TeV are consistent with a quite mild fine-tuning of $\sim$ 10\%. Hence, the upper bounds on stops are neither stringent nor stable under changes of the theoretical scenario. Let us also mention that the fine-tuning problems of the MSSM can be somewhat alleviated by going beyond the minimal scenario, e.g. considering NMSSM or BMSSM frameworks.

\section{Conclusions}\label{}
There is a reasonable hope to find new physics in the Run II of LHC. The naturalness argument supporting the conventional interpretation of the hierarchy problem is still alive, although, for particular BSM scenarios, it starts to be in tension with the (lack of) experimental results. Apart from the hierarchy problem, there are other indications of BSM physics, especially dark matter and flavour physics, which, if we are lucky, could be found at the LHC.  Concerning SUSY, it is still a beautiful candidate for BSM physics, though it seems a bit-fine-tuned. In any case, it is now the time for nature to speak.

\vspace{0.3cm}
{\footnotesize This work has been supported by the MINECO, Spain, under contract FPA2010-17747, FPA2013-44773-P; and by the MINECO Centro de excelencia Severo Ochoa Program under grant SEV-2012-0249. }




\nocite{*}
\bibliographystyle{elsarticle-num}
\bibliography{martin}



\end{document}